\documentclass[reprint,showpacs,preprintnumbers,amsmath,amssymb,superscriptaddress,pre,twocolumn]{revtex4} 
\usepackage[latin1]{inputenc}
\usepackage{amsfonts,amstext}
\usepackage{latexsym}
\usepackage{epic}
\usepackage{graphicx}
\usepackage{array}
\usepackage{color}

\begin{document}

\title{Schelling segregation in an open city: a kinetically constrained Blume-Emery-Griffiths spin-1 system}
\author{Laetitia Gauvin } 
\email{laetitia.gauvin@lps.ens.fr}
\affiliation{Laboratoire de Physique Statistique (LPS, UMR 8550 CNRS, ENS, UPMC Univ. Paris 6 \& Univ. Paris Diderot Paris 7), Ecole Normale Sup\'erieure, Paris, France}
\author{Jean-Pierre Nadal}
\affiliation{Laboratoire de Physique Statistique (LPS, UMR 8550 CNRS, ENS, UPMC Univ. Paris 6 \& Univ. Paris Diderot Paris 7), Ecole Normale Sup\'erieure, Paris, France}
\affiliation{Centre d'Analyse et de Math\'ematique Sociales (CAMS, UMR 8557 CNRS EHESS), Ecole des Hautes Etudes en Sciences Sociales, Paris, France}
\author{Jean Vannimenus}
\affiliation{Laboratoire de Physique Statistique (LPS, UMR 8550 CNRS, ENS, UPMC Univ. Paris 6 \& Univ. Paris Diderot Paris 7), Ecole Normale Sup\'erieure, Paris, France}

\date{\today}

\begin{abstract}
In the 70's Schelling introduced a multi-agent model to describe the segregation dynamics that may occur with individuals having only weak preferences for 'similar' neighbors. 
Recently variants of this model have been discussed, in particular, with emphasis on the links with statistical physics models. Whereas these models consider a {\em fixed} number of agents moving on a lattice, here we present a version allowing for exchanges with an external reservoir of agents. 
The density of agents is controlled by a parameter which can be viewed as measuring the attractiveness of the city-lattice. This model is directly related to the  zero-temperature dynamics of the Blume-Emery-Griffiths (BEG) spin-1 model, with kinetic constraints.
With a varying vacancy density, the dynamics with agents making deterministic decisions leads to a new variety of ``phases'' whose main features are the characteristics of the interfaces between clusters of agents of different types.
The domains of existence of each type of interface are obtained analytically as well as numerically. These interfaces may completely isolate the agents leading to another type of 
segregation as compared to what is observed in the original Schelling model,
and we discuss its possible socio-economic correlates.
\end{abstract}

\maketitle

\section{Introduction}
In the $1970's$, Schelling \cite{Schelling1971,Schelling1978} introduced a model aiming at simulating the interactive dynamics of individuals having specific requirements regarding their neighborhood. This model was based on the intuitive idea that
people of a same type (e.g., White/Black) have some preference for living in a neighborhood where 
the fraction of residents of their own type is not too small. Empirical studies have given some
support to this hypothesis \cite{Clark92,Charles03}, even though such preferences might be the indirect consequence of other factors (e.g., blacks may choose segregated neighborhoods because they have access to social support).
In Schelling's model, agents of two groups with different socio-economic features coexist on a chessboard-city. They individually move to maximize a utility function that depends 
on their tolerance to different neighbors. With such dynamics, a phenomenon of segregation emerges even when the agents are not especially intolerant.
This collective behavior not foreseen in the individual choices has been considered in social science as the paradigm of global phenomena emerging from local social interactions.
Moreover, this model and its outcomes present similarities 
to physical systems as noted , in particular, in \cite{Kirman}, \cite{StS}, \cite{Jensen_competition}. The simplicity of the model combined with its wealth of results, its interdisciplinarity, are obvious
reasons why it has drawn so much attention among scientists.

In the recent years, numerous variants of Schelling's segregation model have been considered. Much as for the Ising model in physics, Schelling's segregation model in social science is the basis for exploring the effect
of various factors on the collective dynamics. 
Among the variants studied, different individual preferences have been dealt with. Schelling initially proposed a binary utility function to summarize the preferences of the agents,
separating their neighborhoods in only two groups: satisfying or not, depending on the fraction of nearby agents of their own type. A further step was to consider other utility functions that may even be continuous. In particular,
it has been shown that segregation may prevail even with agents having a strict preference for a mixed neighborhood \cite{PancsVriend,Zhang}, i.e with an utility 
function which is maximal when there are equal numbers of similar and different neighbors. The influence of the vacancy density on the results has also been widely studied (\cite{CastFL,SiVaWe}).
However, all these models deal with closed systems: a fixed number of agents of each type is allowed to move within the city-lattice.
Schelling himself \cite{Schelling1978} has discussed a case of segregation with an open city,
but within a different setting: he considered a mean-field type model with a city reduced to a single global neighborhood, with a finite number of agents of two types who can only decide either to enter or leave the city, the focus being on the effect of heterogeneity - each agent
having his own tolerance threshold. To our knowledge, there is no study of Schelling's segregation in an open city with local neighborhoods \footnote{The only exception is a study \cite{StS} with a non fixed fraction of agents of each type,  where the Schelling model is simply replaced by the Ising model: this corresponds to a situation with no vacancies and with moves consisting in replacing an agent of a given type by an agent of the other type.}. Adding the possibility for external moves, one has thus a non fixed vacancy density.

Recently, we have shown \cite{LGJVJPN} that Schelling's model is linked to the Blume-Emery-Griffiths (BEG) spin-1 model  with a fixed density of vacancies and with kinetic constraints. The BEG  model \cite{BEG} has been used, in particular, to modelize binary mixtures and alloys in the presence of vacancies. Starting from this correspondence, one can propose a natural generalization of
Schelling's model on an open domain, with a parameter playing the role of a chemical potential for the number of vacancies. The new dynamical system then corresponds to the zero-temperature dynamics \cite{ZTD,ZTD1,ZTD2} of a Blume-Emery-Griffiths model under kinetic constraints. The parameter controlling the vacancy density can be interpreted as measuring the attractivity or hostility of the urban environment. Let us remark that the absence of temperature keeps the model close to the spirit of the original Schelling segregation model which considers deterministic decision processes.

As we will see, a striking feature of this model is to display a variety of interfaces between clusters of agents of a same type. There may be direct contacts between agents of different types as well as lines of vacancies isolating clusters of agents (lines that we henceforth call borders). Such lines have already been observed in some variants of Schelling's model, though in a marginal way since the vacancy density was fixed and therefore could not adjust according to the control parameters to create homogeneous lines.
Here, we exhaustively study the shapes and sizes of these borders and how they are related to the value of the control parameters. Among the relevant points, the absence of contacts between the two types of agents may be considered another kind of segregation, one that is not encompassed by the classical Schelling model.

The paper is organized as follows. In Section \ref{sec:model}, we recall Schelling's original model of segregation and relate it to the BEG model. We then introduce the generalization to the case of an open city. In Section \ref{sec:simul} we present numerical simulations of the model, showing the existence of phases with different types of interfaces. In Section \ref{sec:interfaces} we give the analytical expressions of the phase boundaries in parameter space. Finally in Section \ref{sec:discussion} we discuss the results.

\section{Segregation Model}
\label{sec:model}
\subsection{Contact with the Blume-Emery-Griffiths model}
In Schelling's original model of segregation \cite{Schelling1971}, two types of agents - to be called here and in the following ``red'' and ``blue'' agents- coexist on a regular square lattice with Moore neighborhood (8 next nearest neighbors per site). Each lattice site can be either occupied by a single agent or  vacant. The total number of agents of each type is fixed and kept constant. An agent at a site is said to be satisfied if there is at least a fraction $1-T$ 
of the agents in his neighborhood who are of his own type ($T=2/3$ in~\cite{Schelling1971}), i.e.,
\begin{equation}
 N_d-T(N_d+N_s) \leq 0,
\end{equation}
 where $N_d$ and $N_s$ are ,respectively, the numbers of different and similar neighbors.
The parameter $T$ is called the tolerance. 
Starting with a random configuration, some agents are unsatisfied. With a random sequential dynamics, unsatisfied agents are displaced to a satisfying vacancy (to the closest one in \cite{Schelling1971}).
For a wide range of tolerance $T$ and of vacancy density, the iteration of this process yields regions composed of similar agents (see e.g., \cite{LGJVJPN, Kirman}), phenomenon usually called segregation.

From a physicist's point of view, it is interesting to see that there exists a correspondence  between the Schelling segregation model and spin-1 models.
Let us introduce spin-1 variables $c_i$ taking
the value $0$ if the location $i$ is vacant, and $1$ (resp. $-1$) if this location is occupied by a red (resp. blue) agent. The satisfaction condition at site $i$ can then be written:
  \begin{equation}
     - c_i\, \sum_{\langle j \rangle}c_j -(2T-1) \,c_i^2\sum_{\langle j \rangle}c_j^2 \leq 0,
	\label{eq:dissat_S}
     \end{equation}
where the sums are on the $8$ nearest neighbors of site $i$. Because $c_i=0$ at an empty site, this condition is also true for a site with no agent - one can thus consider the satisfaction condition as a {\em site} property rather than an {\em agent} property.
If an agent is allowed to move from a site where he is not satisfied to an empty site where he is, one can check that the Schelling dynamics admits the following Lyapunov function:
  \begin{equation}
E_S=-\sum_{\langle i,j\rangle} c_ic_j -  K \sum_{\langle i,j\rangle} c_i^2 c_j^2, 
   \label{eq:Es}
  \end{equation}
where $K=2T-1$, and the sum $\sum_{\langle i,j\rangle}$ is on all pairs of nearest neighbors.

This function $E_S$ (\ref{eq:Es}) corresponds to the energy of the Blume-Emery-Griffiths model \cite{BEG}, originally introduced to study the superfluidity of He$^3$-He$^4$ mixtures, under the constraint that the number of sites of each type ($0, \pm 1$) is kept fixed. Hence the Schelling model is equivalent to the 
zero-temperature dynamics of the BEG model, with kinetic constraints (no direct exchange red/blue), and with a fixed number of agents. 
In the full version of the BEG model, the energy contains the additional term $ D_{BEG} \sum_i {c_i}^2 $ (the sum being over all the sites), so that the total number of vacancies is fixed only in average through the Lagrange multiplier $D_{BEG}$:
\begin{equation}
 E_{BEG}= -\sum_{\langle i,j\rangle}c_ic_j -  K \sum_{\langle i,j\rangle} c_i^2 c_j^2 + D_{BEG} \sum_i {c_i}^2.
\label{SI-eq:EBEG}
\end{equation}
 The limit $D_{BEG} \rightarrow  - \infty$ corresponds to the absence of vacancies, i.e., the Ising model, and large positive $D_{BEG}$ corresponds to high vacancy densities. The  term $D_{BEG}$ does not appear in the energy $E_S$ of the Schelling  model, not because it corresponds to $D_{BEG} = 0$, but  because the density of vacancies is fixed. The particular case $K=0$ (that is, $T=1/2$) is known as the Blume-Capel model, much studied for its own sake \cite{Blume,Capel}.

Hence, the obvious next step is to generalize the Schelling model to the case of an open system for which the number of agents is not fixed. In order to allow for exchanges with a reservoir of agents, one has to define the agent's utility in a way allowing to compare
the degree of satisfaction at different locations and between being in or out of the lattice. The full BEG model provides the simplest way to do so, with the parameter $D_{BEG}$ giving  the
 satisfaction loss (if $D_{BEG}<0$) or gain (if $D_{BEG}>0$)  of an agent if he leaves the city from a site with a fully vacant neighborhood. In the next section we detail the resulting multi-agent model.

\subsection{Schelling's model with an open city}
We now specify our variant of Schelling's model for an open system - but still with a lattice of fixed size, the city being not allowed to grow. We introduce an index of dissatisfaction (instead of a ``satisfied or unsatisfied'' binary status). As regards to the dynamics, any agent -whether he is satisfied or unsatisfied- may move to a randomly chosen site if this provides him with a larger degree of satisfaction. As the system is open, we do not fix the total number of agents, nor the ratio of red/blue agents. We assume that there is an infinite reservoir of agents outside the city-lattice.
An agent may leave the city if this increases his degree of satisfaction, and new agents may enter the city. A parameter $D$ controls the flux of agents leaving or entering the lattice, acting as a chemical potential for the vacancy density. 
The control parameters of the model are thus the tolerance $T$ and the vacancy ``chemical potential'' $D$.

Let us now define the index of dissatisfaction.  For an agent within the city (lattice), the index  depends on the heterogeneity of his neighborhood and on an intrinsic attractiveness of the city. The dissatisfaction index $I_{dissat}$ for a neighborhood composed of $N_d$ different and $N_s$ similar neighbors is:
 \begin{equation}
  I_{dissat}= N_d-T(N_d+N_s)+D,
  \label{dissat}
 \end{equation}
where $T$ is the tolerance of the agents about the heterogeneity of their neighborhood. The smaller the index $I_{dissat}$, the more satisfied the agent. With the spin-1 notation
introduced in the preceding section, one can write the dissatisfaction index at site $i$ as
\begin{equation}
  I_i^{dissat}= - \frac {1}{2}c_i\sum_{\langle j \rangle}c_j -\frac {1}{2}(2T-1) \,c_i^2\sum_{\langle j \rangle}c_j^2 + D  \,c_i^2,
  \label{dissat_ci}
 \end{equation}
where, as in (\ref{eq:dissat_S}), the sums are on the $8$ nearest neighbors of site $i$.
Without loss of generality we can assume that an agent outside the lattice has an index of dissatisfaction that is null. This is equivalent to state that $D$ is the satisfaction loss (if $D<0$) or gain (if $D>0$) of an agent who leaves the city from a site with a fully vacant neighborhood - or conversely, it is the gain (if $D<0$) or loss (if $D>0$) of an agent who enters the city at a site with a fully vacant neighborhood.
Hence $D$ indicates how welcoming the lattice is. Indeed, a very negative value of $D$ easily makes the environment satisfying for the agents. While a large value of $D$ leads to the impossibility of satisfying the agents, consequently making the environment hostile. $D$ can be seen as a measure of the (un)attractiveness of the urban environment \cite{Urban_thesis}, and we will call it ``urban attractiveness'' for short (although one should remind that a positive value of $D$ means a hostile environment). In the present socio-economic context, it would have been more suitable to take the opposite sign to define $D$ in Eq.\ref{dissat} but we choose the sign in order to have a direct correspondence with the BEG model.

Starting from a random initial configuration, each agent tries to increase his degree of satisfaction (i.e to decrease $I_{dissat}$). In order to do this, internal or external exchanges are tested with equal probability. In the case of an external exchange, a site is randomly chosen, if it is empty, the  arrival of an agent (one of the two types with equal probability) is attempted. The occupancy becomes actual only if the target dissatisfaction index $I_{dissat}$ is smaller than $0$ (the value of the dissatisfaction index
outside the lattice), that is, if the numbers of similar $N_s$ and different $N_d$ neighbors at the target site satisfy:
 \begin{equation}
  N_d-T(N_d+N_s)+D \leq0.
  \label{utility}
 \end{equation}
If the site is occupied by an agent, the latter remains at this location only if its neighborhood meets the previous condition (Eq.\ref{utility}); otherwise the agent is removed from the lattice. Note that here the tolerance $T$ can be understood as the maximal proportion of different neighbors tolerated in order to remain in the lattice when the environment is neutral (that is, at $D=0$).

In the case of an internal exchange, both a vacant and an occupied site are randomly chosen. The agent moves or not to the empty site  on account of the  difference between the dissatisfaction indexes associated with the two locations: if his current neighborhood is characterized by the set $(N_s,N_d)$ of similar and different neighbors, and the target site by the set $(N_{s'},N_{d'})$, the displacement occurs if: 
\begin{equation}
  [N_{d'}-T(N_{d'}+N_{s'})]-[N_d-T(N_d+N_s)] \leq 0.
  \label{utility2}
 \end{equation}
Let us emphasize that the parameter $D$ disappears in this difference of indexes, so that the internal moves only depend on the tolerance parameter $T$. 

\begin{figure}[h] 
\begin{center}
\includegraphics[width=0.9\linewidth] {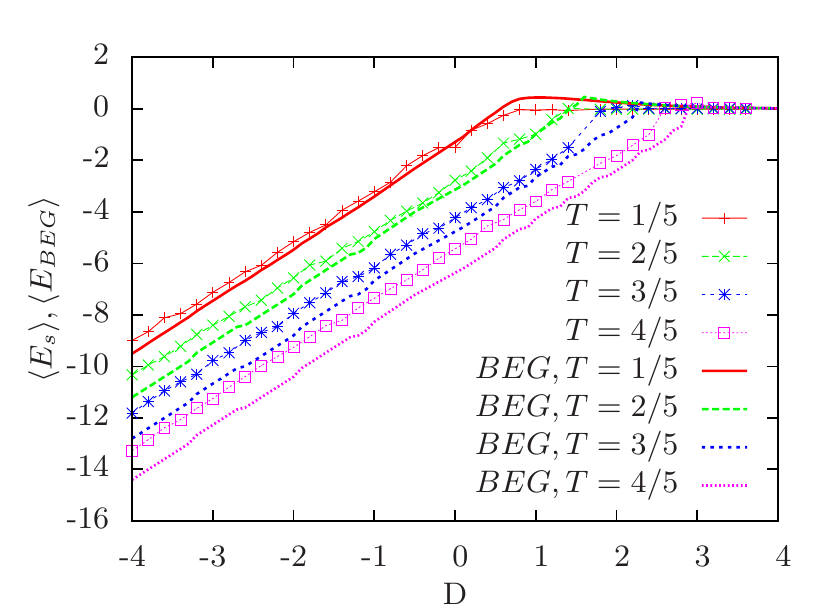}
\caption{(Color online). Mean energy $\langle E_s \rangle$ of Schelling-like  and $\langle E_{BEG}(K=2T-1,D_{BEG}=2D)\rangle$ of Blume-Emery-Griffiths models with respect to $D$ for different values of the tolerance $T$. The quantities have been averaged on $30000$ configurations of $100\times 100$ lattices after equilibrium. As for the Blume-Emery-Griffiths energy, it has been obtained by using a Heat Bath algorithm at a very low temperature.}
\label{enerfig}
\end{center}
\end{figure}
A Lyapunov function of this model is the energy  of the Blume-Emery-Griffiths model (Eq.\ref{SI-eq:EBEG})- with $D_{BEG}$ replaced by $2D$  (see also the expression (\ref{dissat_ci}) of the dissatisfaction index). This shows an equivalence between 
the zero-temperature dynamics of this spin-1 model and the present model, except for the following kinetic restrictions: here red agents cannot become blue and conversely. The energy is always decreasing during the dynamics, but the constrained dynamics create 
barriers between local minima. Consequently, the energy does not necessarily reach its absolute minimum.
Figure \ref{enerfig} shows the dependency on $D$ of the energy at the fixed point (or in the stationary regime), for different fixed values of the tolerance $T$. On the same figure, the energy of the corresponding BEG model at zero temperature limit is also shown: one observes only weak differences between the two models.
We attribute this weak difference to the fact that 
the absence of thermal noise makes the interfaces robust, the latter contributing to surface terms in the energy which are small compared to the dominant volume terms.

\section{Numerical simulations}
\label{sec:simul}
\begin{figure}[h]	
\begin{center}
\includegraphics[width=8cm]{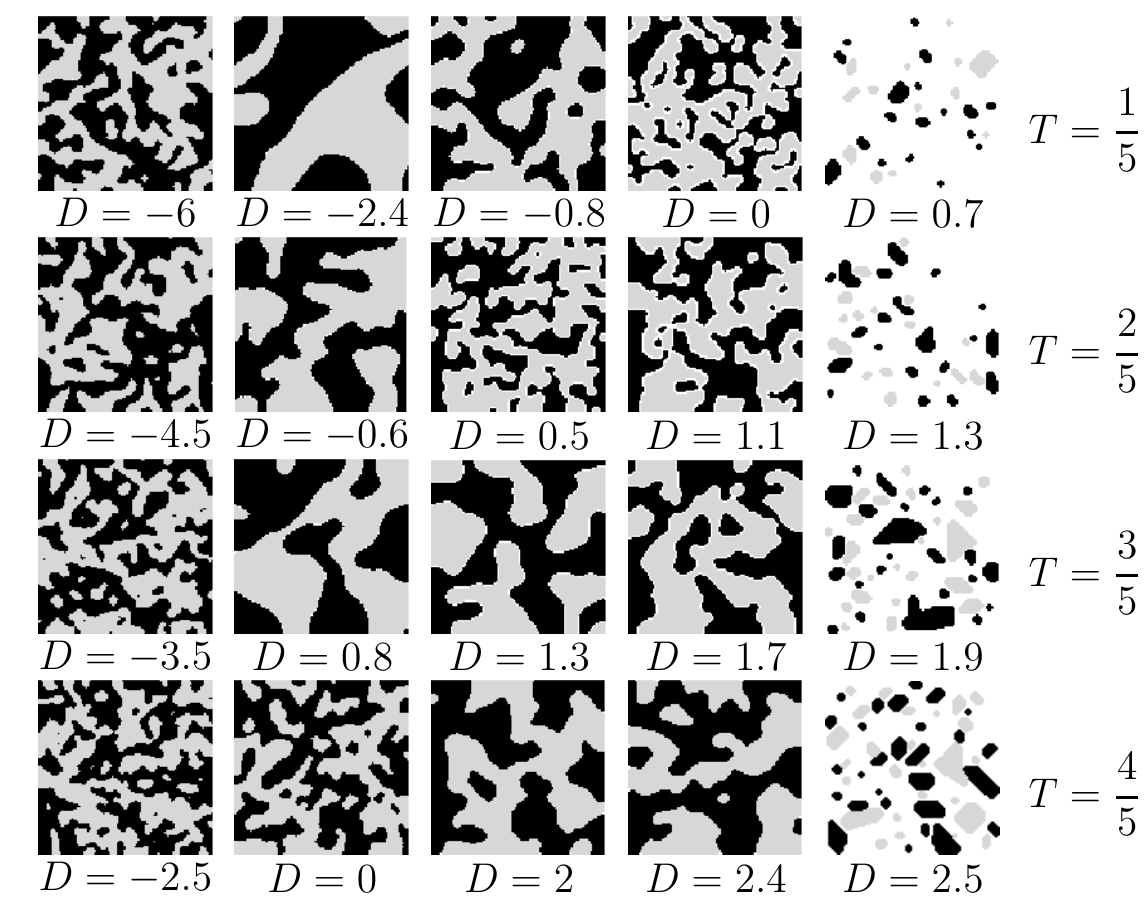}
\caption{Final configurations obtained for different values of the parameters $T$ and $D$ with $100\times 100$ lattice. Dark and  light gray pixels correspond to the two types of agents whereas white pixels represent vacant sites.}
\label{cf}
\end{center}
\end{figure}

Numerical simulations are performed on a $L\times L$ lattice ($L=100$) with free boundary conditions for different values of the tolerance $T$ and of urban attractiveness $D$. The initial configurations are fully mixed as the agents and vacancies are randomly placed on the lattice. The dynamics previously described is applied until the system reaches equilibrium, i.e, when the computed quantities only have very weak fluctuations or when the system is frozen. Figure \ref{cf} shows the final configuration for different values of $T$ and $D$.

The variety of observable configurations for the different values of the urban attractiveness $D$ depends on the tolerance $T$. However, there are two extreme phases that we meet at each tolerance: one without vacancies and one dominated by vacancies. Actually, at highly negative $D$, for any tolerance, no vacancy is present. The agents of the two types are in direct contact. Indeed, the lattice is so welcoming that agents prefer to remain in it even with different neighbors. Conversely, for the high values of $D$, the environment is so unwelcoming that agents massively leave. In-between, according to the value of the tolerance considered, vacancies may appear and create interfaces isolating the two types of agents. Thus, the interfaces between red and blue agents are of several types: the contact can be direct or vacancies can separate the agents. The interfaces also have several shapes: rugged or smooth giving more or less compact clusters. A quantitative analysis will give the existence domains of these interfaces.

\subsection{Numerical Analysis: Types of interfaces}
\label{sec:numeric-analysis}
\begin{figure}[h] 
\begin{center}
\includegraphics[width=0.9\linewidth] {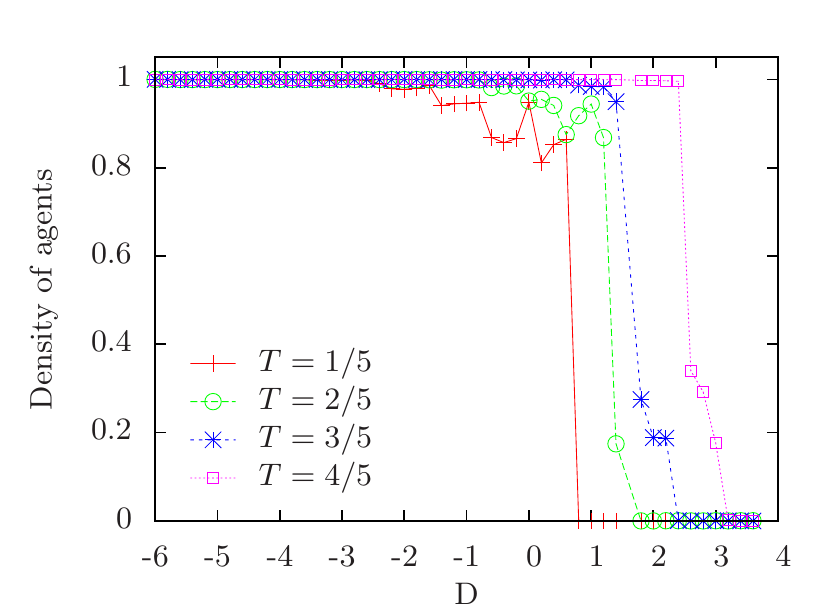}
\caption{(Color online).
Density of agents (total number of agents divided by the size of the lattice $L \times L$)  versus the urban attractiveness $D$ for different values of the tolerance $T$ computed on $L\times L$ $(L=100)$ lattices. The number of agents is here an average over $30000$ configurations after equilibrium has been reached.}
\label{Nagt}
\end{center}
\end{figure}

To study the occurrence of the phases, we compute different quantities such as the mean numbers of different neighbors per site and the density of agents. The number of agents allows to locate the transition to the state of vacancies. This quantity has been plotted on Fig.\ref{Nagt} versus the urban attractiveness for several values of the tolerance.

The density of agents first slightly departs from its maximum value. This corresponds to the appearance of the first vacancies. As $D$ increases, the density of agents abruptly falls to a very low value showing a discontinuous transition to the predominant vacancy state. This decrease does not lead to a zero value of the density because some small clusters of agents remain in a ``sea of vacancies''. Note that for high $T$, the appearance of the first vacancies almost coincide with the sharp transition to the predominant vacancy state.

\begin{figure}[h]
\begin{center}
\includegraphics[width=0.9\linewidth]{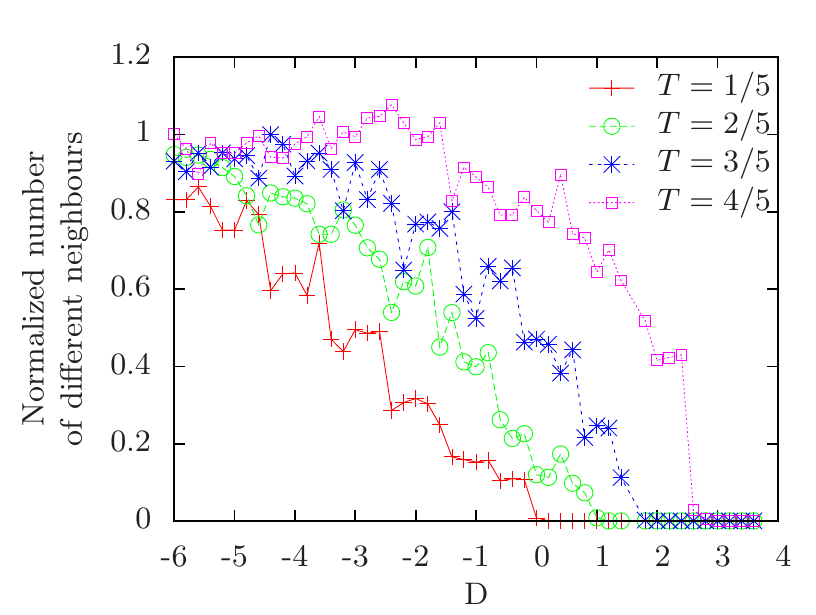}
\caption{(Color online). Mean normalized number of different neighbors versus the urban attractiveness $D$ for different values of the tolerance $T$ computed on  $100\times 100$ lattices: the number of different pairs of neighbors has been averaged over $30000$ configurations after equilibrium.}
\label{Nd}
\end{center}
\end{figure}

The evolution of the number of different neighbors (Fig.\ref{Nd}) yields the outline of the interface type. As $D$ is increasing, the number of links between the two different types of agent decrease as a result of either the growth of the clusters or the departure of different neighbors.

\begin{figure}[h]		
\begin{center}
\includegraphics[width=4.5cm] {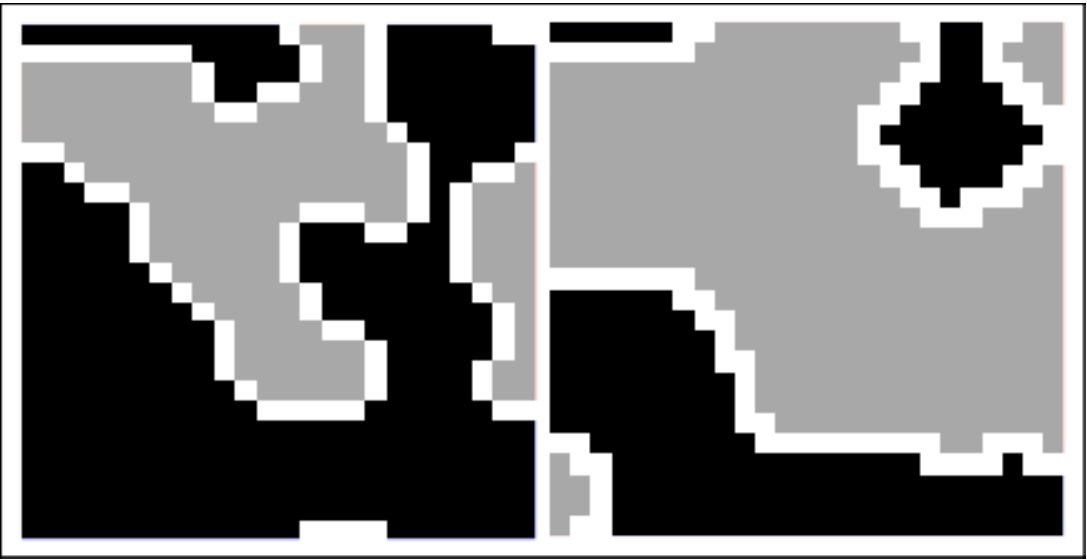}
\caption{Types of vacancies interface: width-1 (left) and width-2 (right) interfaces.}
\label{front_vac}
\end{center}
\end{figure}

When the number of different neighbors becomes null, a full vacancy interface separates the clusters. There no longer exists contact between different agents. Let us remark that before a full vacancy interface appears when $D$ is increased, a thiner interface of vacancies is observed.  This interface that we will characterize as width-1, is such that a path can be followed  between agents by going only ``through''  neighboring vacant sites without encountering an occupied site. However diagonal contacts between different agents may still exist. This is not the case for the  full vacancy interface that we will call width-2. Figure \ref{front_vac} illustrates both types of interfaces by zooming in two configurations.

The plots of the previous quantities on Fig.\ref{Nagt} and Fig.\ref{Nd} give the existence domains of the different types of interfaces. However they do not provide much information about the shapes of the interfaces.

\subsection{Numerical Analysis: Shapes of interfaces}
\begin{figure}[h]	
\begin{center}
\includegraphics[width=0.9\linewidth] {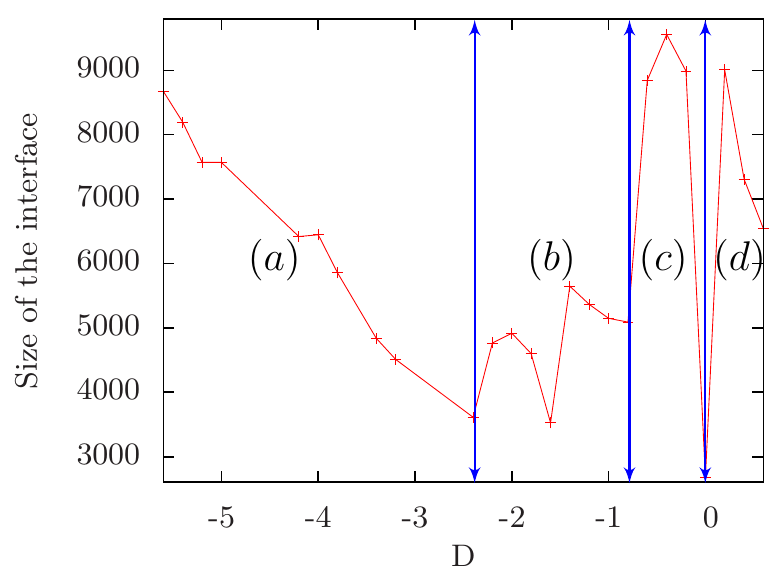}
\includegraphics[width=0.7\linewidth] {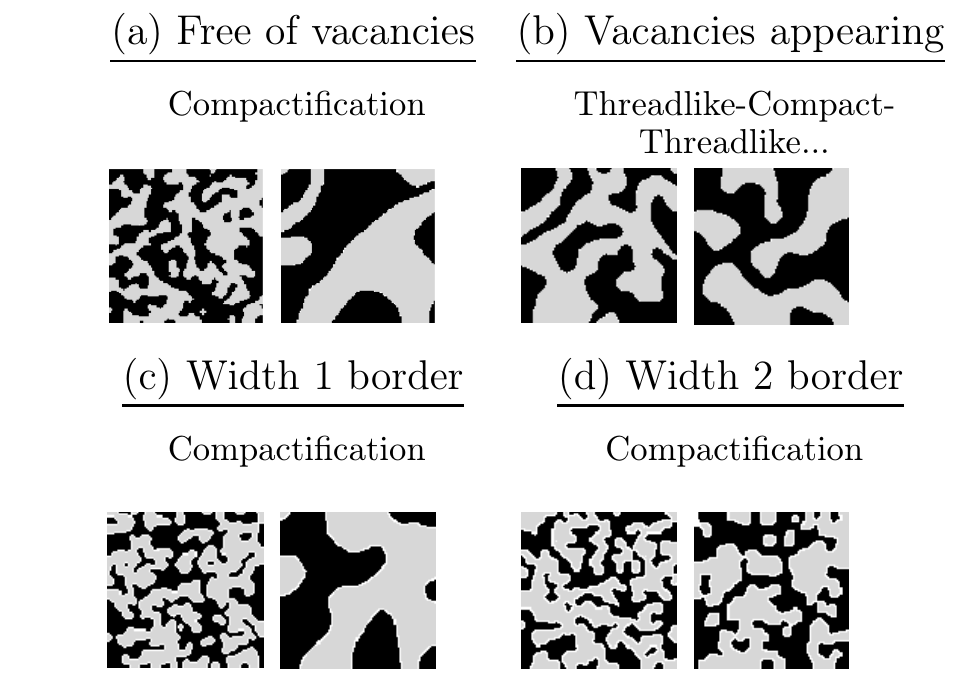}
\caption{Size of the interface for $T=1/5$. The same patterns and same order of magnitude are observed with different initial conditions.  Below the curve, we present the typical equilibrium configurations observed for each phase (free of vacancies, full vacancy border $\dots$) at the smallest and largest $D$ values of the phase. In each phase, the clusters observed in the equilibrium configurations eventuate to become more compact as $D$ increases.}
\label{DP}
\end{center}
\end{figure}

To determine the evolution of the shape of the interfaces between agents, we measure the size of the interface. This measure, plotted in Fig.\ref{DP}, is the sum of the numbers of different neighbors and the number of contacts between vacancies and agents.

In the phase free of vacancies, the size of the interface decreases when increasing $D$, meaning that the clusters become larger and larger to limit the number of contacts between red and blue agents.
As for the phases whose equilibrium configurations comprise width-2 vacancy interface (resp. width-1 vacancy interface), when $D$ increases within the domain of existence of the phase considered, the size of the interface first increases then decreases. This behaviour can be explained as follows: at the lowest values of $D$ for which the considered regime exists, the interface is rugged, the agents accept vacancies (resp. different agents and vacancies) in their neighborhood instead of similar neighbors. But as $D$ increases, the environment becomes less welcoming. As a consequence, the agents reorganize themselves in compact clusters so as to increase the number of similar agents with respect to the number of vacancies (resp. different agents and vacancies) in their neighborhood. This reorganization renders the interfaces smooth and yields a decrease in total energy even with a large value of $D$.

\subsection{Remarks}

Let us notice that the main results do not depend on the initial conditions chosen in the simulations.  The results presented on all the figures have been obtained with random initial configurations, 
red agents, blue agents and vacancies being uniformly distributed in the same proportions ($1/3$) on the lattice.
If, when increasing $D$, instead of drawing a new random configuration for each new value of $D$, we take as initial configuration the equilibrium configuration obtained at the previous value of $D$, the existence domains of the phases are not modified. Differences does occur, but only concern the evolution with $D$ of the size of the clusters:starting from the previous equilibrium configuration leads to steadily growing clusters with increasing $D$, which is not the case when starting each simulation with a new random initial configuration as described above. 
When using the same protocol but with a decreasing $D$, the locations of the transitions are roughly the same as those obtained by starting from random initial configurations, except for the transition to the predominant vacancy state. Indeed, given that 
the latter state is stable for any positive value of $D$, the transition only occurs at $D=0$. Basically, unless we take extreme initial configurations (for example, vacancies state), the existence domains of the phases do 
not depend on the initial configuration.
Differences can only be seen in the shapes of the interfaces. Indeed, if we begin with an 
equilibrium configuration obtained for a set of values $(D,T)$ where the agents are less easily satisfied,
and therefore already presents some structure, it necessarily leads to more compact clusters than when starting with a random initial configuration.

\section{Theoretical Analysis: Interfaces}
\label{sec:interfaces}

By studying the moves allowed by the dynamics, one can predict the type of interface for ranges of $T$ and $D$.

\subsection{State free of vacancies}
 Let us take a configuration free of vacancies and write the conditions for the appearance of at least one vacancy. The vacancies appear only if at least one agent has an unsatisfying neighborhood, i.e., the dissatisfaction index $I_{dissat}$ (Eq.\ref{dissat}) is positive. Since here $N_s+N_d=8$, this condition can be written as follows:
 \begin{eqnarray}
  N_s &\leq& D+8(1-T).
  \label{vac1}
 \end{eqnarray}
 This resulting vacancy will not be occupied by an agent of the other type only if the number of different neighbors satisfy this inequality as well:
 \begin{eqnarray}
  N_d &\leq& D+8(1-T).
  \label{vac2}
 \end{eqnarray}
 Therefore, adding the two previous inequalities, one gets that the vacancies appear only if:
 \begin{eqnarray}
 D\geq-4+8T.
 \label{vac}
 \end{eqnarray}

\subsection{Vacancy interface}
Let us assume that the system reaches a configuration with a complete vacancy interface separating the different clusters, i.e., no contact between different agents exists. For at least one agent (let us call it $A$) to tolerate one different neighbor, the number of identical agents $N_s$ around him must verify the inequality:
 \begin{eqnarray}
  1-T(N_s+1)+D &\leq& 0,\\
 TN_s &\geq& 1-T+D.
 \end{eqnarray}
The number of similar neighbors $N'_s$ of his single different neighbor (let us say $A'$) has to satisfy this inequality as well:
\begin{eqnarray}
 TN'_s\geq 1-T+D.
\end{eqnarray}
 Both agents $A$ and $A'$ have at least two neighbors in common and they are neighbors with one another, which leads to:
\begin{eqnarray}
 N_s&\leq& 5,\\
 N'_s&\leq& 5.
\end{eqnarray}
Moreover, at least two neighbors of $A$ are in the neighborhood of two neighbors of $A'$. As we assume there is no pair of different agents, this constrains two of them to be vacancies. The sum of the similar agents of $A$ and of $A'$ must be lower than eight. The first link between red and blue agent appears only if:
\begin{eqnarray}
8T \geq TN_s + T  N'_s &\geq & 2(1-T+D),\\
D &\leq 5T-1. 
\end{eqnarray}
Consequently, if a complete vacancy interface (no contact between different agents) exists, it subsists at least for $D \geq 5T-1$.
The same kind of argument can be made to show that an agent can have two different neighbors only if:
\begin{eqnarray}
D &\leq 6T-2. 
 \end{eqnarray}
In other words, if there exists a vacancy interface of width 1, i.e such that we can travel all the way  by following only nearest-neighbors (Moore neighborhood) vacancies and which allows diagonal contacts between agents, this vacancy interface subsists at least for $D \geq 6T-2$.

\subsection{Predominant vacancy state}
 The agents do not leave the lattice if the following condition is fulfilled:
 \begin{eqnarray}
  N_d\leq \frac{-D + TN_s}{1-T}.
  \label{vacst}
 \end{eqnarray}
It requires a number of similar neighbors at least as great as $D/T$. Let us notice that the agents and vacancies are initially uniformly distributed. As a result, a large majority of agents do not have more than three similar neighbors. If $D/T>3$, the agent will leave in massive numbers and will never come again because almost no agent have more than three similar neighbors. Consequently, if $D/T>3$, the vacancies are predominant in the equilibrium state.

\subsection{Summary: Domains of existence of interfaces}
\begin{figure}[ht]	
\begin{center}
\includegraphics[width=8cm] {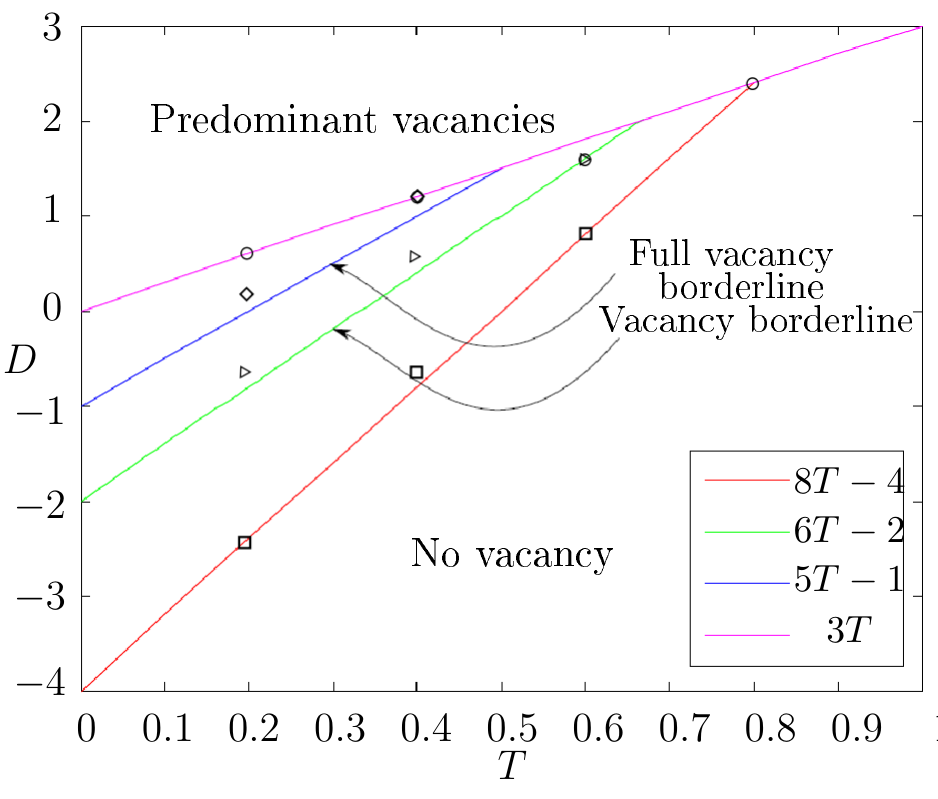}
\caption{(Color online). ``Phase diagram'' of our model in the parameter space (attractiveness $D$, tolerance $T$).
Theoretical phase diagram: lines determined analytically (see text), limiting the domains of existence of the different states -- free of vacancies, with predominant vacancy and with vacancy interface. Simulation points are marked on the diagram: the squares, triangles and diamonds indicate the points at which, respectively, the first vacancies, the width-1 interface and the width-2 interface appear. The circles corresponds to the last points at which the vacancies are not predominant. Remark: the simulations have been done for values of $D$ incremented by a constant factor $0.2$: this gives the numerical precision of the location of the phases.)}  
\label{DP2}
\end{center}
\end{figure}

 These analyzes are summarized on the phase diagram represented on Fig.\ref{DP2},
giving the domains of existence of the different types of interfaces.

The theoretical phase boundaries, obtained from studying the dynamical stability of the interfaces,
are in good agreement with what we found via the numerical simulations presented above, section \ref{sec:numeric-analysis}, as illustrated by the simulations points marked on the figure \ref{DP2}.

\section{Discussion}
\label{sec:discussion}
We have introduced a variant of the multi-agent segregation model of Schelling in which agents are allowed to leave or enter the ``city''. The dynamics of this model admits a Lyapunov function which makes it identical to the zero-temperature dynamics of the  Blume-Emery-Griffiths model with kinetic restrictions. In this model, phases with new features compared to both the BEG and Schelling's models emerge.

It is worthwile discussing with more details the correspondence with the  Blume-Emery-Griffiths model.
Let us first briefly recall the nature of the phases of the latter. Several studies (\cite{BEG,BEG1,BEG2}) shows the presence of ferromagnetic, paramagnetic, predominant vacancies and anti-quadrupolar phases whose typical configurations are lattices with two sublattices, one with spins equal to zero, one with alternated spins $-1,+1$. These configurations do not appear for the range of parameters for which we have a correspondence with our variant of the Schelling model. Moreover at low temperature and in particular in the  zero temperature limit, in the BEG model only ordered phases -- ferromagnetic or vacancies phases -- may exist in the domain of parameters that we are interested in.

In the Schelling-type open model discussed here, a phase of vacancies appears at $D=3T$ instead of $D=2(K+1)=4T$ (corresponding to $D_{BEG}=8T$) in the Blume-Emery-Griffiths model. Actually, since the direct exchange between agents of different types is not allowed by the dynamics, agents do not have other choices than to leave the system to become satisfied. That is why, the equilibrium configurations with a high density of vacancies
appear earlier in the multi-agent model. In the range of values of the parameters at which the ferromagnetic phase exists in the BEG model, the deterministic dynamics with kinetic constraints gives rise to a stable state with red and blue agents coexisting. Indeed, in this regime, the agents reorganize themselves to reduce their dissatisfaction index but they do not need to leave the lattice to be satisfied. Consequently, both types of agents are present on the lattice and as the system is kinetically constrained, the completely red or blue configurations do not appear even if this would minimize the total energy. Moreover, such coexistence can be sustained even with moderately friendly environments (not too negative $D$), thanks to the adjustment of the interfaces.

As said above, for the range of parameters that we consider, the ground state of the Blume-Emery-Griffiths model is either ferromagnetically ordered or ordered with all spins equal to zero. The BEG ground states may show clustering with vacancy interfaces (which can be flat or rough) but only
if $J+K$, the sum of the  bilinear and biquadratic interaction coefficients, is negative 
\cite{Rough-Transition,Ground-States}. Since in our model $J$ is equal to $1$, we have always $J+K=2T>0$. This shows that the interfaces appearing in our study do not correspond to configurations giving the absolute minima of the BEG energy, but result from the constrained dynamics.

At this point one may ask what would be the results if a thermal noise was introduced in our model while keeping the same kinetic constraints. One may consider the dynamics where the randomly chosen agent decide to move or not (to a vacant site or outside the city) according to a Glauber type choice rule with a temperature like parameter $1/\beta$: the move is accepted if it is favourable to the agent, and otherwise accepted with probability $\exp{(-\beta \Delta I_{dissat})}$ where $\Delta I_{dissat}$ is the increase in dissatisfaction index. One can then show that, despite the kinetic constraints forbid a direct exchange between agents of different types, the system reaches an equilibrium distribution given by the Gibbs distribution of a BEG model at temperature $1/\beta$, with an effective value of $D$ which is $D_{eff}=2D+\frac{\ln2}{\beta}\;$ \footnote{The proof is left to the reader. One shows that the considered dynamics is a Monte Carlo dynamics for a BEG model at the temperature $1/\beta$, with a transition matrix which is irreducible and satisfies the detailed-balance condition. If $q_{\pm}$ is the fraction of agents of type $\pm 1$ in the reservoir, for any non zero rate of exchanges with the reservoir one gets as stationary state the Gibbs distribution associated to the energy $E_{BEG}(K=2T-1,2D) -\frac{1}{2 \beta} [ \ln (q_+ q_-) \sum_i c_i^2  +  \ln \frac{q_+}{q_-} \sum_i c_i ]$. In the present case $q_{\pm}=1/2$.}. At a small value of the temperature $1/\beta$, one will have a long transient during which  interfaces separating clusters of agents of different type will appear. The time scale on which the segregation of the zero-temperature dynamics will appear and be sustained will be larger if one increases the rate at which external exchanges are tried.

It is interesting to note that, in contrast with the BEG model, the dynamics in the present model leads to segregation, but does not lead to an ordered state which would mean the full exclusion of one type of agent.

Let us now come back to Schelling's model and more generally to the socio-economic context.
In Schelling's original model and its variants which consider a fixed number of agents and vacancies, 
it is known that a phenomenon of segregation appears. This segregation is defined as the grouping of agents of the same type together, while the vacancies have no functional role. They do not display patterns and thus play no major role in the process of segregation. Indeed, as shown for the variant introduced in \cite{LGJVJPN},
the vacancies are uniformly distributed in the equilibrium configurations whatever the phase, segregated, mixed or mixed-frozen.  There is no regard on the interfaces between the groups. In the open model studied here, a completely different kind of segregation consisting in the isolation of groups of agents from each other by vacancies is exhibited. This phenomenon can be linked to the percolation of vacancies in the network whereas, with a fixed vacancy density, what matters is the percolation of the agents of a given type \cite{LGJVJPN}.

With internal/external exchanges being assimilated to, respectively, moving in/out a given area like a city, it may not seem realistic to have the same probability for both types of exchanges. However, internal exchanges only induce a minimal level of clustering and thus have only a limited impact on the existence domains of the phases. The equilibrium configurations are conditioned by the external exchanges. Let us notice that if we use a weak rate of external exhanges, the system converges to the final states after a long transitory phase while the internal exchanges dominate the dynamics. As a result, the proportion of internal exchanges mainly influences the shapes and size of the clusters but not the type of the interfaces.

To conclude, we recall and stress that, in the present work, the build-up of interfaces through the appearance of vacancies is the main focus. Depending on the tolerance threshold $T$ and the city (un)attractiveness $D$, the vacancies may be sporadically distributed along the interface or form complete connected vacancy borders (large or thin) between agents. The equilibrium configurations highlight the permanent competition between searching a neighborhood providing a high enough satisfaction with respect to the level of attractiveness of the environment, and leaving the city altogether. Contrarily to what would be obtained with a thermal noise, the possibility to leave the city does not lead to an unstability which would make the city occupied by agents of a same single type. Here both types of agents coexist in the city, but the dynamics leads
to a segregation into clusters with a variety of interfaces between the clusters. We classify these interfaces according to two features:
\begin{itemize}
 \item their type: blue and red agents may be in contact or separated by vacancies,
 \item their shape: rugged or smooth.
\end{itemize}
As a matter of fact, the most important conclusion is that here vacancies have a functional role, they allow weakly tolerant agents to be satisfied. This is not the case in Schelling's original model where the vacancies are only ``conveyor of moves''. The functional character of the vacancies is clearly established with the formation of interfaces.  
When vacancies do not single-handedly allow the agents to be satisfied by decreasing their number of different neighbors, the interfaces become smooth to increase the number of similar ones.

Finally, one may speculate on the interpretations of the results in a socio-economic context. 
The compactification of clusters of similar agents encountered when the environment becomes hostile (large positive $D$) may be reminiscent of the strengthening of the links between people via a community network that sociologists have observed, for example, in some neighborhoods of Chicago \cite{Neighborhood}. These strong links may prevent a massive exodus due to the lack of attractiveness of the environment.
As for the presence of full vacancy interfaces obtained for some range of values of the control parameters, they separate groups in a way which reminds of 
socio-spatial segregation reinforced by walls, as in Johannesburg \cite{Frontier}. In the model, we observe the formation of frontiers with a homogeneous network with no infrastructure or other physical boundaries. Clearly preexisting structural borders (roads, parks, rivers...) may affect segregation (Ex: Paris ring road \cite{Frontier}). How would these physical borders affect our results? In the range of parameters where vacancy borders appear, the segregation dynamics can be expected to take advantage of such pre-existing frontiers which would then become parts of the social borders. In other regimes, even if the physical frontiers may facilitate the emergence of an interface, they will not necessarily lead to social frontiers. It would therefore be interesting to perform empirical studies in order to see under which conditions a physical border is at the same time a social border.

More generally, further studies should focus on variants that take into account realistic socio-economic features of the agents or infrastructures of the city. For example, one could categorize agents according to their income, associate prices to the vacant sites and move the agents according to their financial capacities. Also, the level of attractiveness of the urban environment $D$ could be heterogeneous on the lattice in order to reflect the presence of more or less facilities in different areas. In the same vein, one could make the attractiveness depend on the type of agent who looks at a vacant place to account for various subjective criteria, and agents could be given an idiosyncratic tolerance threshold. It would be interesting to model the presence of ethnical minorities by changing the relative proportion of agents in the reservoir, that is, by modulating the proportion of attempted arrivals of each type of agent. Finally, we have considered an open system with an urban domain of a fixed size: one could deal with a completely open system by letting free the size of the urban system. This would address issues related to the urban sprawl, of different natures than the one related to segregation.

\subsection*{Acknowledgments} 
We thank the referees for very useful comments. LG thanks RB for helpful discussions. LG is supported by a fellowship from the French Minist\`ere de l'Enseignement Sup\'erieur et de la Recherche allocated by the UPMC doctoral school ``ED389: Physics, from Particles to Condensed Matter''. JPN and JV are CNRS members.
This work is part of the project ``DyXi'' supported by
the  program SYSCOMM of the French National Research Agency (grant ANR-08-SYSC-008).

\end{document}